\documentclass[amsmath,twocolumn,prl,superscriptaddress]{revtex4}
\usepackage{graphicx}
\usepackage{dcolumn}
\usepackage{bm}
\usepackage{amsmath,amssymb}
\usepackage{ulem}
\usepackage{color}
\usepackage{amsfonts}
\usepackage{bigints}

\begin{document}

\title{Thermoelectric response across the semiconductor-semimetal transition \\in black phosphorus}

\author{Yuna Nakajima}
\affiliation{Department of Physics, Gakushuin University, Tokyo 171-8588, Japan.}

\author{Yuichi Akahama}
\affiliation{Graduate School of Material Science, University of Hyogo, Kamigori 678-1297, Japan.}

\author{Yo Machida}
\email{yo.machida@gakushuin.ac.jp}
\affiliation{Department of Physics, Gakushuin University, Tokyo 171-8588, Japan.}

\date{\today}

\begin{abstract}
In spite of intensive studies on thermoelectricity in metals, little is known about thermoelectric response in semiconductors at low temperature. An even more fascinating and unanswered question is what happens to the Seebeck coefficient when the semiconductor turns to a metal. By precisely tuning the ground state of black phosphorus with pressure from the semiconducting to semimetallic state, we track a systematic evolution of the Seebeck coefficient. Thanks to a manifest correlation between the Seebeck coefficient and resistivity, the Seebeck response in each conduction regime, i.e., intrinsic, saturation, extrinsic, and variable range hopping (VRH) regimes, is identified. In the former two regimes, the Seebeck coefficient behaves in accordance with the present theories, whereas in the later two regimes available theories do not give a satisfactory account for its response.
However, by eliminating the extrinsic sample dependence in the resistivity $\rho$ and Seebeck coefficient $S$, the Peltier conductivity $\alpha=S/\rho$ allows to unveil the intrinsic thermoelectric response, revealing vanishing fate for $\alpha$ in the VRH regime. 
The emerged ionized impurity scattering on entry to the semimetallic state is easily surpassed by electron-electron scattering due to squeezing of screening length accompanied by an increase of carrier density with pressure.
Each carrier scattering participates an enhancement of the phonon drag contribution to the Seebeck effect, but creates the phonon drag peak with opposite sign at distinct temperature.  
In the low temperature limit, a small number of carriers enhances a prefactor of $T$-linear Seebeck coefficient as large as what is observed in prototypical semimetals. A crucial but largely ignored role of carrier scattering in determining the magnitude and sign of the Seebeck coefficient is indicated by the observation that a sign reversal of the $T$-linear prefactor is concomitant with a change in dominant scattering mechanism for carriers.
\end{abstract}

\maketitle

\section*{I. INTRODUCTION}
When a solid is subject to temperature gradient, thermal diffusion of charge carriers 
develops electric field along the trajectory of carriers.
This is a phenomenon know as the Seebeck effect.
The magnitude of this effect is quantified by Seebeck coefficient $S$, which is defined as a ratio of temperature difference $\Delta T$ to thermoelectric voltage $V_{\rm th}$, $S=-V_{\rm th}/\Delta T$.
Let us begin by showing that depending on what statistics electrons obey the Seebeck coefficient behaves differently between metals and semiconductors.
In the Boltzmann picture with relaxation time approximation, the Seebeck coefficient is expressed by
\begin{equation}
S=-\cfrac{1}{eT}\left\{\dfrac{\bigintss_{0}^{\infty}\dfrac{\partial f_0}{\partial\epsilon}D(\epsilon)\epsilon(\epsilon-\mu)\tau(\epsilon) d\epsilon}{\bigintss_{0}^{\infty}\cfrac{\partial f_0}{\partial\epsilon}D(\epsilon)\epsilon\tau(\epsilon) d\epsilon}\right\},
\end{equation}
where $f_0$ is quilibrium distribution function and
the density of state is $D(\epsilon)=(2m)^{3/2}\epsilon^{1/2}/2\pi^2\hbar^{3}$.
The scattering time is assumed to be $\tau(\epsilon)=\tau_0\epsilon^{r}$.
By using the Fermi integral $F_n(\eta)=\int^{\infty}_{0}f_0(\zeta,\eta)\zeta^{n}d\zeta$,
Eq.~(1) is rewritten as~\cite{ioffe}
\begin{equation}
S=\cfrac{k_{\rm B}}{e}\left\{\eta-\cfrac{\left(r+\cfrac{5}{2}\right)F_{r+3/2}(\eta)}{\left(r+\cfrac{3}{2}\right)F_{r+1/2}(\eta)}\right\},
\end{equation}
where $\zeta=\epsilon/k_{\rm B}T$ and $\eta=\mu/k_{\rm B}T$ ($\mu$ being the chemical potential).

For metals the Fermi-Drac distribution function applies to $f_0$.
Employing the Sommerfeld expansion for large positive values of $\mu/k_{\rm B}T$, the Seebeck coefficient for electrons in metals is given by
\begin{equation}
S=-\frac{\pi^2}{3}\frac{k_{\rm B}}{e}\frac{k_{\rm B}T}{\epsilon_{\rm F}}\biggl(r+\frac{3}{2}\biggr).
\end{equation}
Here, $\epsilon_{\rm F}$ is the Fermi energy and we assume $\mu=\epsilon_{\rm F}$.
From Eq.~(3) one finds that the magnitude of Seebeck coefficient is set by the Fermi energy $\epsilon_{\rm F}$ and the exponent of energy dependence of scattering time $r$, which takes different values depending on the scattering mechanism: $r=-1/2$ for phonon scattering, $r=0$ for neutral impurity scattering, and $r=3/2$ for ionized impurity scattering~\cite{ioffe,seeger}. $S$ linearly decreases with temperature and vanishes at zero temperature.
Validity of this expression is documented by a universal relation between $S/T$ and the Sommerfled value $\gamma$ as well as $S/T$ and the prefactor of $T^2$ resistivity for a wide variety of metals~\cite{behnia,okazaki}.

For insulators the Maxwell-Boltzmann distribution function applies to $f_0$, then the Seebeck coefficient is obtained to be
\begin{equation}
S_{\rm e,h}=\mp\frac{k_{\rm B}}{e}\biggl\{\frac{\Delta}{2k_{\rm B}T}+\biggl(r+\frac{5}{2}\biggr)\biggr\},
\end{equation}
where $\Delta$ is a gap between the valence and conduction band.
The suffixes e and h denote electron and hole, respectively.
One can see that the magnitude of Seebeck coefficient is set by $\Delta$ and $r$, and it diverges on cooling inversely proportional to temperature.
This expression well describes the Seebeck coefficient of semiconductors in the intrinsic regime as reported in Ref.~\cite{johnson}.

Here several questions arise.
Q1 What is the fate of Seebeck coefficient in semiconductors even below the intrinsic regime?
Surprisingly, after 200 years of discovery of Seebeck effect there is no consensus about what is going on in this fundamental transport quantity when a semiconductor is cooled down to low temperatures.
The limited number of studies have reported the contradicting observations; it diverges in some cases~\cite{liuSi,machidaPF} and vanishes as $T\rightarrow0$ in other cases~\cite{buhannic,demishev,ishida,berthebaud}.
Q2 What does happen in the Seebeck coefficient at the metal-insulator (M-I) transition? In other words, how do the contrasting Seebeck responses between metals and insulators converge at the transition?
The thermoelectricity near the M-I transition has been barely explored~\cite{lakner,ishida}. The pioneer work has done by H. v. Lohneysen {\it et. al.}~\cite{lakner} on the doped Si, 
who found that the Kondo effect emerges close to the critical doping. Because of this extrinsic contribution induced by impurities, the systematic evolution of the intrinsic diffusive contribution to Seebeck coefficient 
near the M-I transition is hardly observable.
Q3 How does the electron scattering affect the Seebeck coefficient?
Since the magnitude of Seebeck coefficient is primaliry detrmined either by $\epsilon_{\rm F}$ or $\Delta$,
in most cases much attension has not given to the scattering parameter $r$ in Eqs. (3) and (4). 

Black phosphorus (BP), an elemental semiconductor~\cite{ling}, provides a rare opportunity to address the above issues. 
Prior transport studies have revealed the resistivity behavior like the doped (p-type) semiconductor despite that BP is undoped~\cite{akahama}.
The atomic vacancy is attributed to the hole doping~\cite{riffle}.
The low density of carriers ($\sim$10$^{15}$ cm$^{-3}$) leads to a heat transport dominated by phonons, which yield a variaty of fascinating phenomena including 
phonon hydrodynamics~\cite{machidabp} and sizable phonon Hall effect~\cite{li}.
Application of hydrostatic pressure exceeding 1 GPa is enough to close the band gap of $\sim$0.3 eV and the semimetallic state is stabilized~\cite{akiba1,xiang,akiba2}.
It should be emphasize that the semiconductor-semimetal transition is isostructural, thus it is the Lifshitz transition~\cite{xiang}.
This is in contrast with the pressure-induced M-I transition in Si which is accompanied by a stractural change~\cite{minomura}.
The ability to use the pressure to tune the ground state is advantage of BP over such as Si~\cite{lakner} because an introduction of impurities can be avoided.
The Dirac nature of mobile carries created in the semimetallic state makes this elemental semimetal more attractive~\cite{xiang,gong,fujii}.

Our focus is to unveil the thermoelectric property of semiconductor and its evolution on entering the semimetallic state across the semiconductor-semimetal (S-S) transition utilizing BP as a representative semiconductor.
Our precise transport measurements under pressure up to 1.7 GPa and in the temperature range between 2 K and 300 K
allow to track the systematic evolution of resistivity and Seebeck coefficient as a function of temperature and pressure.
We show that thanks to the clear correspondence to the resistivity which can be separated into four conduction regimes, i.e., intrinsic, saturation, extrinsic, and variable range hopping (VRH) regimes, the characteristic Seebeck response in each regime is resolved.
In particular, the vanishing fate for Seebeck coefficient is uncovered in the VRH regime.
While the Seebeck behavior in the intrinsic and saturation regimes is satisfactory accounted by the present theories, the responses in the extrinsic and VRH regimes are not in agreement with any available theories.
However, we show that being free from extrinsic sample dependence, the Peltier conductivity arrows to access the intrinsic thermoelectric response at low temperature.

Concomitant with an entry to the semimetallic state, ionized impurity scattering becomes dominant scattering process affecting electrons, but it is overwhelmed by electron-electron scattering due to shrinkage of screening length of impurity potential upon increasing of density of conduction electrons with pressure.
We show that both scattering could be sources to enhance the phonon drag contribution to the Seebeck coefficient.
A direct link between the change in hierarchy of dominant scattering and a sign reversal of the $T$-linear prefactor of Seebeck coefficient indicates that a crucial role is played by carrier scattering in determining the magnitude and sign of the Seebeck coefficient,
although its importance has been largely ignored. 

\section*{II. METHODS}
Single crystals of black phosphorus were synthesized under high pressure~\cite{endo}.
Seebeck coefficient under pressure was measured by the differential DC method using 
chromel-constantan thermocouples as probes of thermoelectric voltage. Schematic image of Seebeck coefficient measurement setup is shown in Fig.~1. The junctions of the thermocouples were made by spot welding, and glued directly to the sample by using silver paste. A temperature gradient across the sample
was generated by a chip resistor mounted on the thin glass-epoxy flame. The resistor was connected to the hot end of the sample by a gold wire which was also used as a current electrode for the resistivity measurements. The cold end of the sample was thermally anchored at a large copper plate with silver paste to establish steady heat flow from the sample.
The absolute Seebeck coefficient $S$ of the sample is evaluated from the following relation,
\begin{equation}
S=\frac{S_{\rm ch}-(V_{\rm ch}/V_{\rm co})S_{\rm co}}{1-(V_{\rm ch}/V_{\rm co})},
\end{equation}
where $V_{\rm ch}$ ($V_{\rm co}$) is voltage across the circuit consisting of the sample and chromel (constantan), and $S_{\rm ch}$ ($S_{\rm co}$) is an absolute Seebeck coefficient of chromel (constantan).
$S_{\rm co}$ was evaluated from the differential Seebeck coefficient of the type E thermocouple with independently measured $S_{\rm ch}$~\cite{slack,zrudsky,chaussy}.
Effect of pressure on the type E thermocouple was found to be negligible (see APPENDIX).

External pressure was applied by using a NiCrAl-BeCu hybrid piston cylinder. 
Daphne oil 7373 was used as a pressure transmitting medium.
The applied pressure was estimated by monitoring the superconducting transition temperature of Pb.
The resistivity was also measured under the same pressure environment utilizing the voltage electrodes for the thermoelectric measurements.

\begin{figure}[tb]
\begin{center}
\centering
\includegraphics[width=8cm]{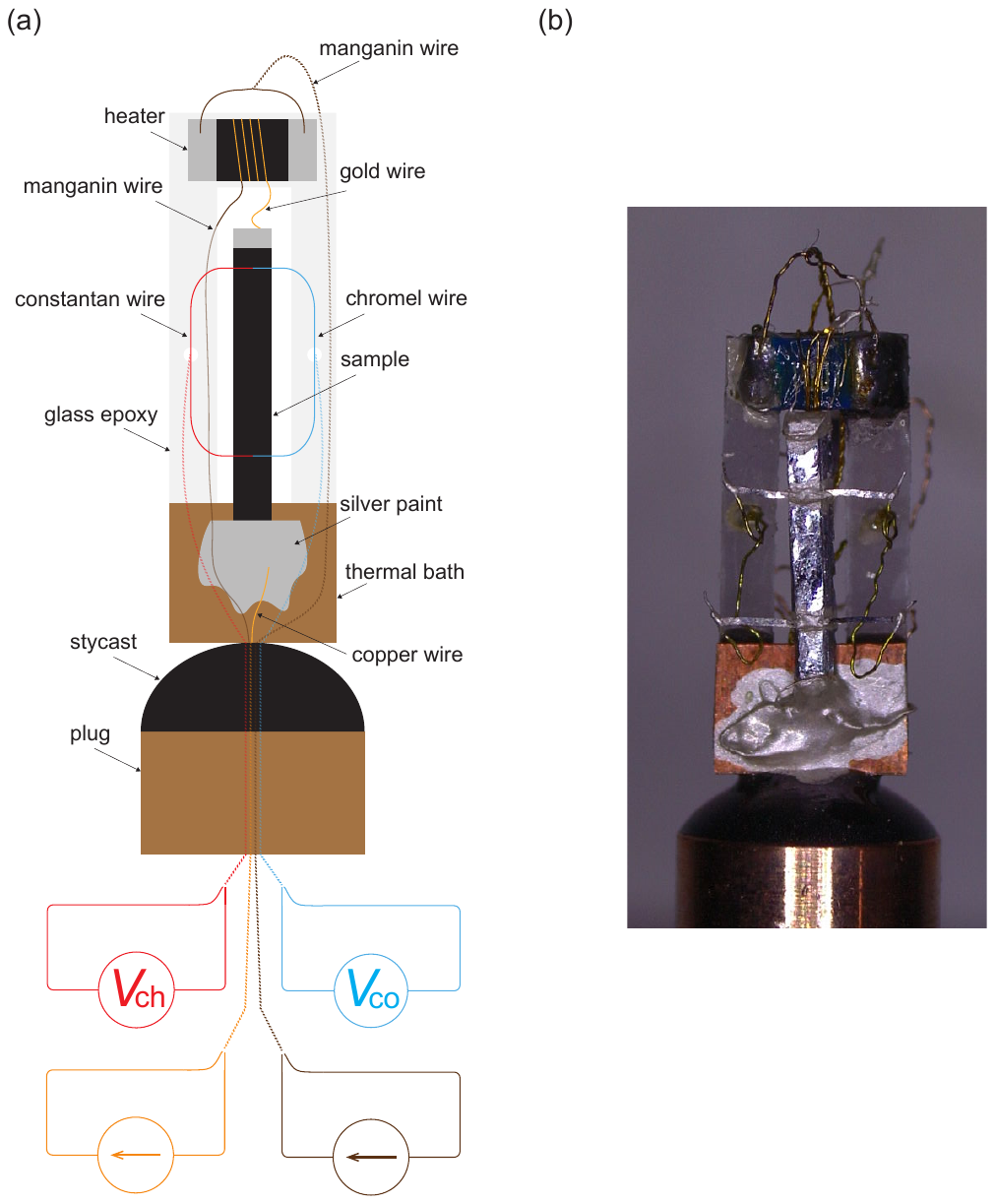}
\vspace*{-3cm} 
\caption{(a) Schematic of Seebeck coefficient measurement setup. 
(b)A  photo of the setup.}
\end{center}
\end{figure}

\section*{II. RESULTS}
\subsection*{Electrical resistivity}
Figure~2 shows temperature dependence of electrical resistivity $\rho(T)$ measured under various pressures.
$\rho(T)$ at ambient pressure can be separated into four conduction regimes, i.e., intrinsic, saturation, extrinsic, and variable range hopping (VRH) regimes.
In the intrinsic regime between 250 K and 300 K, $\rho$ shows an activation behavior. By fitting the data to the Arrhenius equation $\rho=\exp(\Delta_{\rm vc}/2k_{\rm B}T)$ (Fig. 3(a)), we obtained an activation energy of $\Delta_{\rm vc}\sim$ 0.3 eV which corresponds to an energy gap between valence and conduction bands.
The estimated $\Delta_{\rm vc}$ is in agreement with what was reported in the previous study~\cite{keyes}.
$\rho(T)$ decreases with temperature in the saturation regime between 50 K and 250 K, reflecting the facts that the carrier density is insensitive to the temperature and that a scattering rate of electrons by phonons is reduced on cooling. 
In the extrinsic regime between 15 K and 50 K, $\rho(T)$ again exhibits the activation behavior (Fig. 3(b)) with an energy gap of $\Delta_{\rm va}\sim$16 meV between the valence band and acceptor level.
Below 15 K, an evolution of $\rho$ becomes weak due to an entrance in the VRH regime where the conduction is governed by a hopping of electrons between impurity sites. In this regime, $\rho$ varies following the expression $\rho\propto\exp[(T/T_{\rm 0})^{-1/(d+1)}]$  ($d$ is a space dimension). As seen in Fig.~3(c)-(e), our limited data does not allow to determine the hopping dimensionality.

\begin{figure}[tb]
\includegraphics[width=8cm]{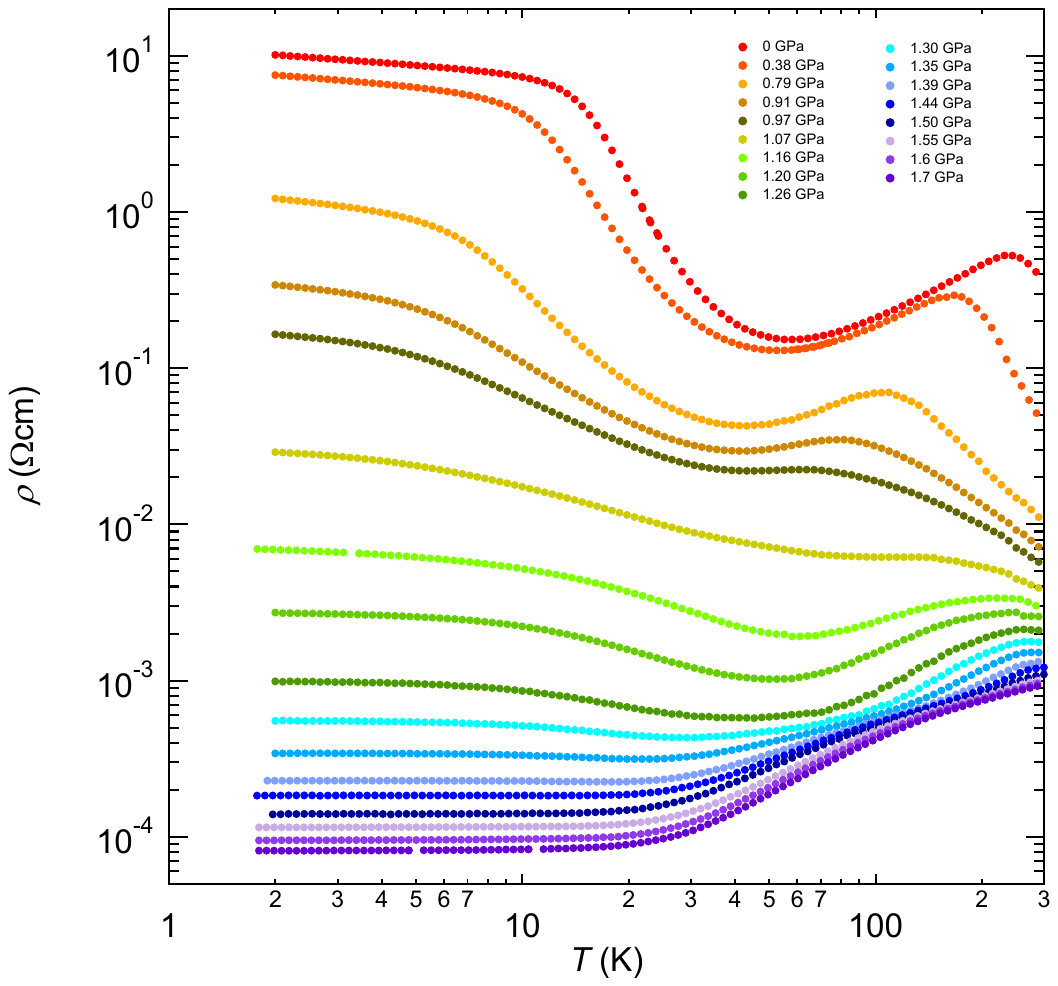}
\vspace*{-5cm} 
\caption{Temperature dependence of resistivity $\rho$ under various pressure in a logarithmic scale.}
\end{figure}
\begin{figure}[tb]
\includegraphics[width=9cm]{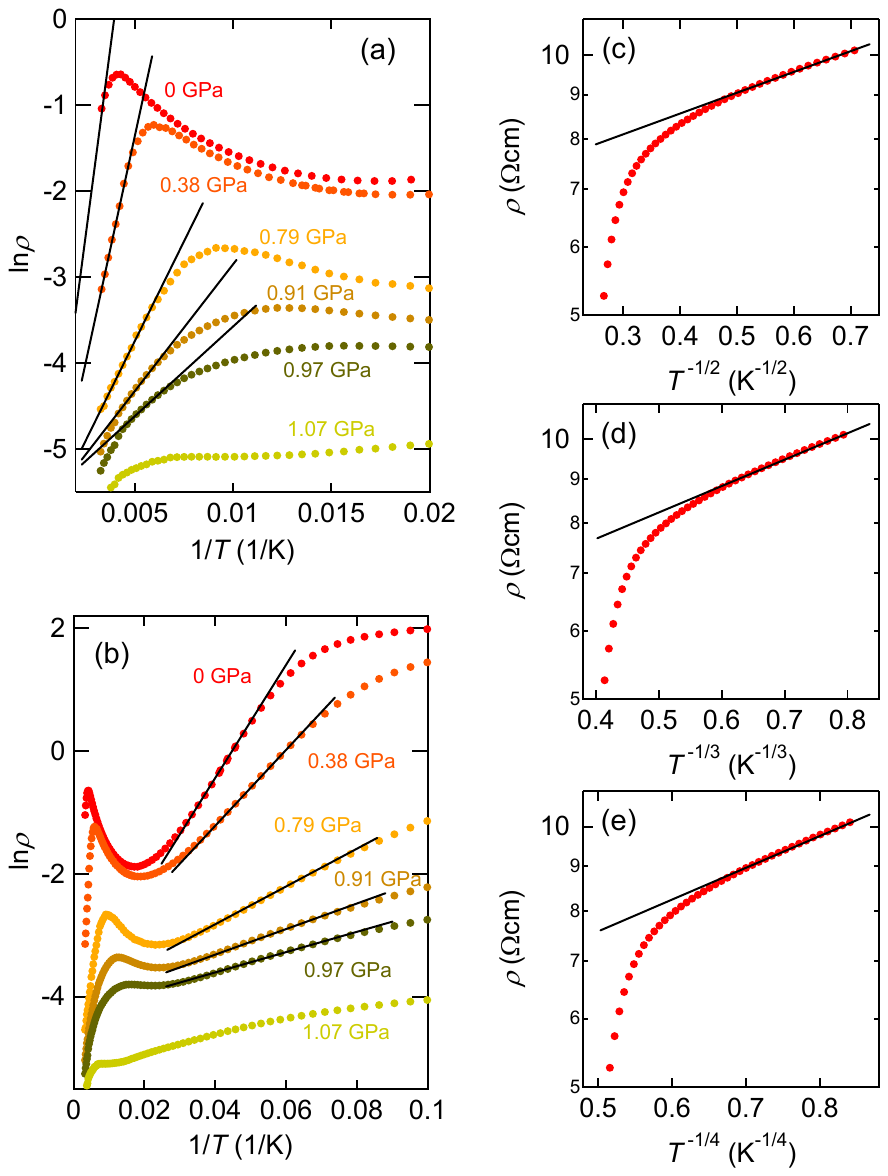}
\vspace*{-4cm} 
\caption{Arrhenius plots in the intrinsic and extrinsic regimes are shown in the panel (a) and (b), respectively. Resistivity in the VRH regime is plotted against (c) $T^{-1/2}$, (d) $T^{-1/3}$, and (e) $T^{-1/4}$.}
\end{figure}

By increasing the pressure, a rise of $\rho$ in the intrinsic and extrinsic regimes becomes weak due to the gap closing.
Indeed, the gap estimated from the Arrhenius plot (Figs. 3(a) and (b)) decrease with pressure as shown in Fig. 5(a). A linear extrapolation suggests that the gaps vanish around $P_{\rm c}\sim$ 1.1 GPa. 
This conjecture is supported by the observation that the semiconducting behavior in $\rho(T)$ becomes unclear at $P$ =1.07 GPa (Fig. 2(a)).

\begin{figure}[tb]
\includegraphics[width=7cm]{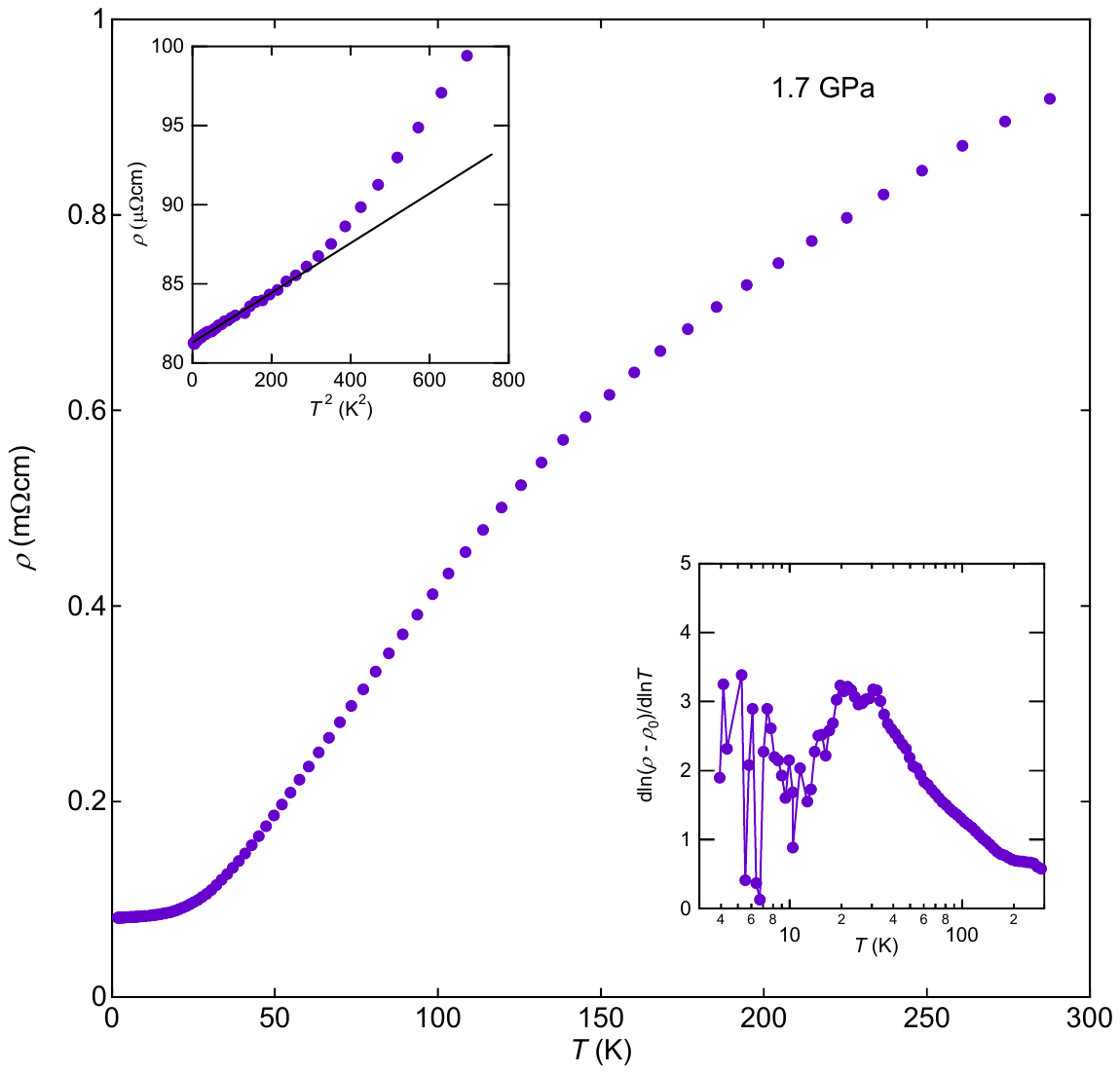}
\vspace*{-3.5cm} 
\caption{(a) Temperature dependence of resistivity $\rho$ under the pressure of 1.7 GPa. 
At low temperature, $\rho$ follows $T^2$ as shown in the upper inset. Lower inset: exponent $\gamma$ of temperature dependence of resistivity ($\rho=\rho_0+A_{\gamma}T^{\gamma}$) which is quantified by taking the
logarithmic derivative after subtracting residual resistivity: $\gamma=\{d\ln (\rho-\rho_0)/d\ln T\}$ is shown as a function of temperature.}
\end{figure}

Above 1.16 GPa, a metallic behavior eventually shows up in $\rho(T)$ at high temperature, but below 100 K $\rho(T)$ subsequently rises. Such an upturn is observed in gapless semiconductors and semimetals~\cite{blatt,carvalho,goswami,jaouiInAs}, and is attributed to the ionized impurity scattering. 
For such a scattering process, screening of the impurity potential by conduction electrons is irrelevant.
A comparable magnitude of the Thomas-Fermi screening length $r_{\rm TF}\propto\sqrt{a_{\rm B}/n^{1/3}}\sim$ 15~nm and the mean-free-path $l$ = 24~nm calculated from resistivity $\rho=3\pi^2\hbar/e^2k_{\rm F}^2l$~\cite{gunnarsson}
implies the ionized impurity scattering is marginal in BP.

In fact, by further increasing the pressure, the upturn entirely suppressed and $\rho$ keeps decreasing all the way down to the lowest temperature at the pressure exceeding $P^{*}\sim$ 1.35 GPa. 
This is caused by the squeezing of $r_{\rm TF}$ upon increasing the number of conduction electrons, and thus the scattering from ionized impurity becomes ineffective.
In Figure 4, a plot of $\rho$ vs $T$ at $P$ = 1.7 GPa is shown as representative temperature dependence of $\rho(T)$ in the semimetallic state. As seen in the upper inset of Fig. 4, at low temperature $\rho(T)$ exhibits $T^2$ dependence arising from electron-electron scattering. A linear fit to the data gives a $T^2$ prefactor of 15.2~n$\Omega$cm/K$^2$, which is comparable to the one find in Bi~\cite{hartman}.
On warming, electron-phonon scattering becomes dominant and as a result, an exponent of $T$ dependence of $\rho$ increases from two to three as demonstrated in the lower inset of Fig. 4 where the logarithmic derivative after subtracting residual resistivity:
dln$(\rho-\rho_0)$/dln$T$ is plotted as a function of temperature. By further increasing temperature, the exponent drops to approach around one. Similar variation of dln$(\rho-\rho_0)$/dln$T$ has been observed in Sb~\cite{jaoui}

\begin{figure}[tb]
\includegraphics[width=9cm]{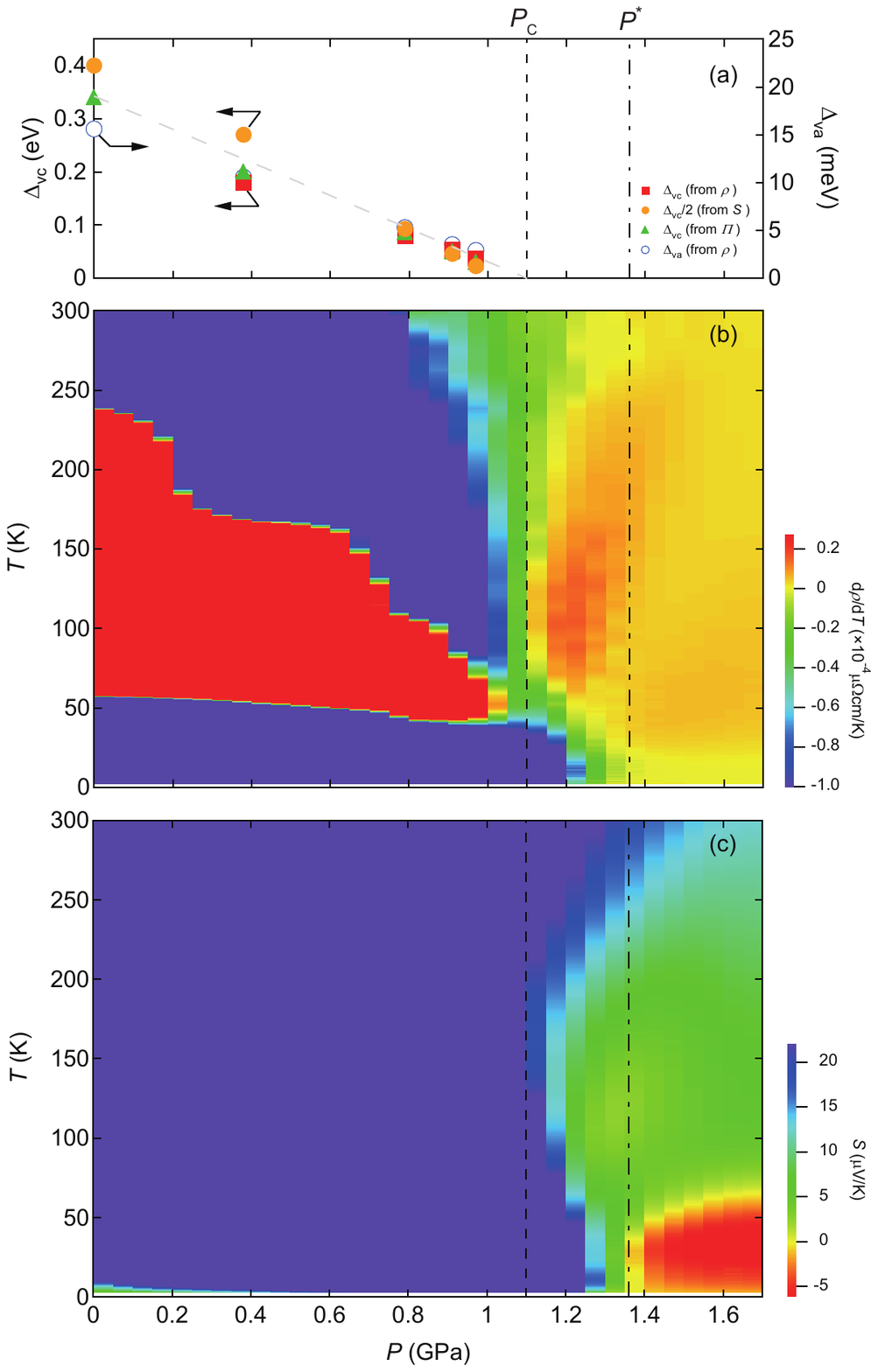}
\vspace*{-0.5cm} 
\caption{(a) The gaps between the top of valence band and the bottom of conduction band $\Delta_{\rm vc}$ (left axis) determined by resistivity and Peltier coefficient decrease linearly with pressure, pointing that the gap vanishes around 1.1 GPa as denoted by a dashed line. The gap between the top of valence band and the acceptor level $\Delta_{\rm va}$ (right axis) follows the same trend. Contour plot of temperature derivative of resistivity and the Seebeck coefficient in the temperature-pressure plane are shown in the panels (b) and (c), respectively.}
\end{figure}

One can see how $\rho$ evolves under pressure from a contour plot of temperature derivative of $\rho$, d$\rho$/d$T$, in the temperature-pressure plane, which is shown in Fig. 5(b).
In the plot, warm/cold color represents a region where $\rho$ decrease/increase on cooling.
It is clearly seen that the intrinsic, saturation, and extrinsic/VRH regions colored by purple, red, and purple, respectively, are squeezed upon increasing the pressure and the former two regions are terminated at the S-S transition ($P_{\rm c}\sim$ 1.1 GPa) denoted by a dashed line.
While the extrinsic/VRH region also disappears at the same pressure, the emergence of ionized impurity scattering above $P_{\rm c}$ extends the cold-color region (d$\rho$/d$T<0$) to the semimetallic state. 
As can be seen by a change of color from clod to hot across $P^{*}\sim$ 1.35 GPa (a dashed-and-dotted line), the dominant scattering mechanism switch from the ionized impurity scattering to electron-electron scattering.
As we will show below, this change in scattering mechanism inside the semimetallic state provides a great impact on the Seebeck effect.

\subsection*{Seebeck coefficient}
\subsubsection*{semiconducting state}
Figure 6 depicts the temperature dependence of Seebeck coefficient $S(T)$ under various pressures.
A sign of $S$ is positive in the semiconducting state, consistent with that black phosphorus is the p-type semiconductor.
At ambient pressure, $S(T)$ increases with decreasing temperature and attains a maximum value, 800 $\mu$V/K around 200 K.  
Up to the critical pressure ($P_{\rm c}\sim$ 1.1 GPa), the maximal $S$ is diminished by a factor of five.
The magnitude of $S$ keeps decreasing even above $P_{\rm c}$ as displayed in Fig. 6(b),
and $S$ takes place a sign change at $P^{*}\sim$ 1.35 GPa (the inset of Fig. 6(b)) at low temperature. The contour plot of $S$ displayed in Fig.~5(c) clearly shows that the sign change in $S$ is concomitant with the disappearance of ionized impurity scattering. 
 
\begin{figure}[tb]
\includegraphics[width=7cm]{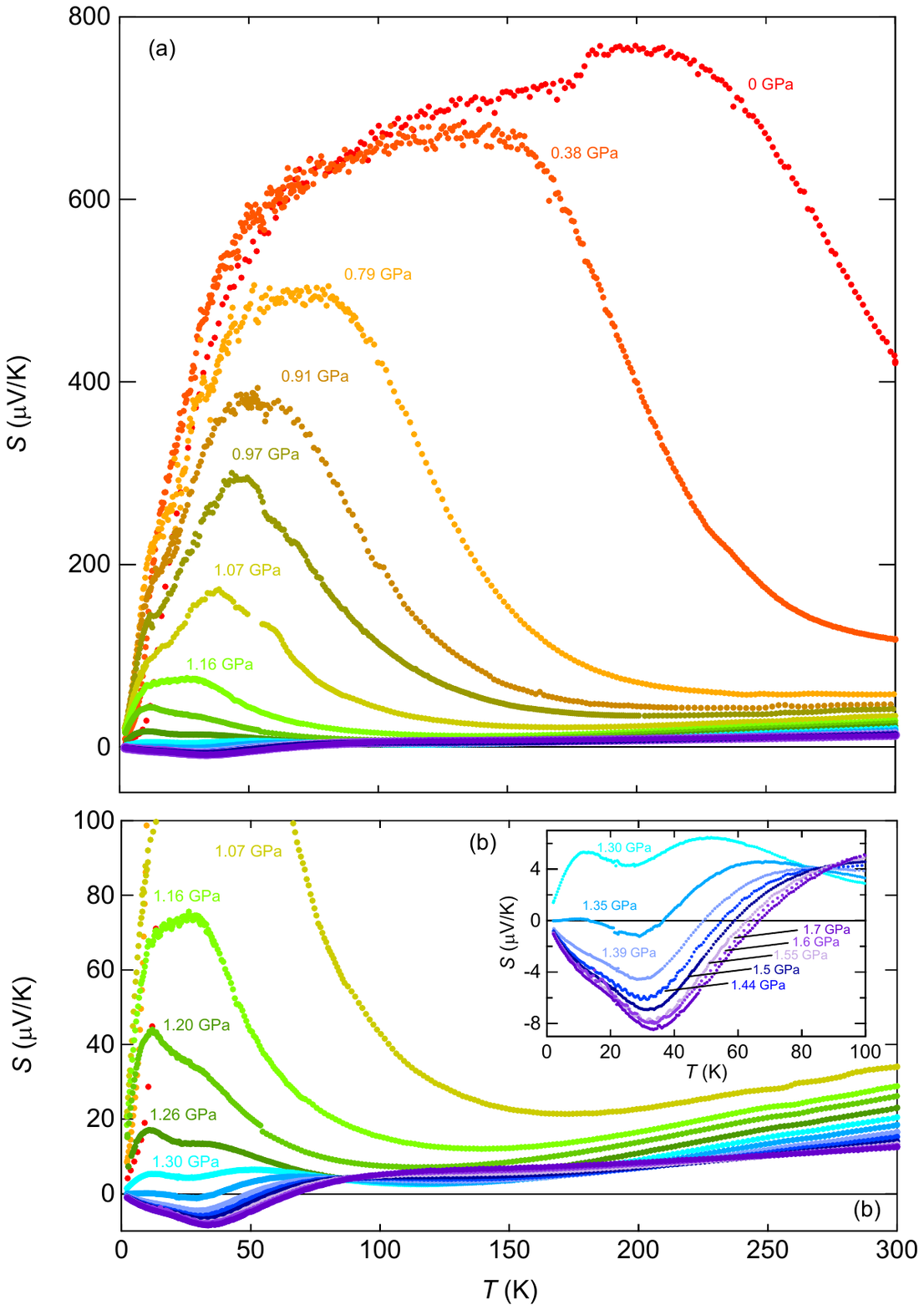}
\vspace*{-0cm} 
\caption{(a) Temperature dependence of Seebeck coefficient $S$ under various pressure. Corresponding zooms above 1.07~GPa and 1.30~GPa are shown in the panel (b) and its inset, respectively.}
\end{figure}
\begin{figure*}[htb]
\includegraphics[width=18cm]{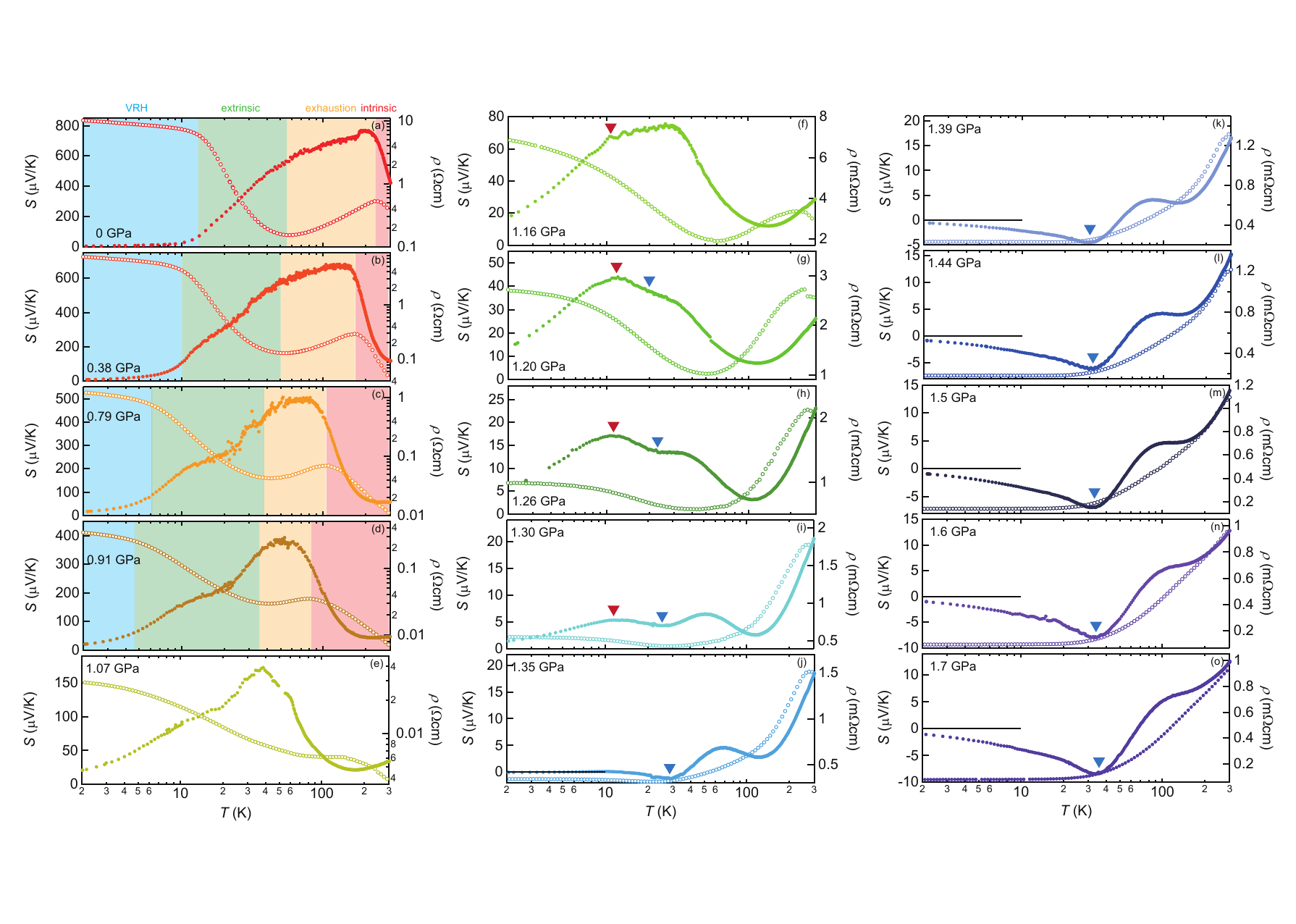}
\vspace*{-1.5cm} 
\caption{Temperature dependence of Seebeck coefficient (left axis) together with resistivity (right axis) under each pressure. Panels (a)-(d) and (e)-(o) correspond to the data in the semiconducting and semimetallic states, respectively. In the panels (a)-(d), four conduction regimes in the semiconducting state are highlighted by different colors.}
\end{figure*}

One of the main outcome of this study is a manifest correlation between resistivity and Seebeck coefficient.
Figure 7 shows the temperature dependence of Seebeck coefficient (left axis) together with resistivity (right axis) at each pressure.
In the semiconducting state, Seebeck coefficient shows characteristic behaviors in four conduction regimes: $S$ sharply rises in the intrinsic regime, then becomes nearly temperature independent in the saturation regime, and subsequently drops rapidly in the extrinsic regime. Finally, $S$ again  decreases slowly in the VRH regime.
An application of the pressure makes the intrinsic and extrinsic regime wider and the saturation and VRH regime narrower, creating a prominent peak of $S$ in the saturation regime.
On entering the semimetallic state at $P>P_{\rm c}$, the peak in $S$ becomes less pronounced.

From Eq. (4), the diffusion thermopower is expected to evolves with $\Delta_{\rm vc}/T$ in the intrinsic regime. Indeed, this is seen in our data as shown in Fig. ~8(a) where $S$ is plotted against $1/T$. 
An extraction of $\Delta_{\rm vc}$ from the slope is not straightforward due to the presence of thermally excited hole and electron carriers.
The Seebeck coefficient of two carrier system is expressed as
\begin{equation}
S=\frac{\sigma_{\rm e}S_{\rm e}+\sigma_{\rm h}S_{\rm h}}{\sigma_{\rm e}+\sigma_{\rm h}},
\end{equation}
where $\sigma_{\rm e}$ ($\sigma_{\rm h}$) and $S_{\rm e}$ ($S_{\rm h}$) are conductivity and Seebeck coefficient of electrons (holes), respectively.
By substituting Eq.~(4) to Eq.~(6), we get
\begin{equation}
S=\cfrac{k_{\rm B}}{e}\cfrac{1-(\sigma_{\rm e}/\sigma_{\rm h})}{1+(\sigma_{\rm e}/\sigma_{\rm h})}\biggl\{\frac{\Delta}{2k_{\rm B}T}+\biggl(r+\frac{5}{2}\biggr)\biggr\}.
\end{equation}
Suppose that $\sigma_{\rm e}/\sigma_{\rm h}\sim1$, the Seebeck coefficient will be very small when the chemical potential is located close to middle of the energy gap. As the chemical potential moves toward one of the bands (the valence band for the case of BP) on cooling, 
$\sigma_{\rm e}/\sigma_{\rm h}$ becomes small and the Seebeck coefficient is dominated by the contribution from holes.
This is what occurs in our sample when it is cooled from room temperature.
By assuming that the Seebeck coefficient is dominated only by holes, $\Delta_{\rm vc}$ is obtained by fitting the data to Eq.~(4).
As seen in Fig.~5(a), the extracted $\Delta_{\rm vc}$ is about twice as large as the one from resistivity.
This discrepancy may be due to the additional contribution by electrons and/or the presence of two kinds of holes with distinct mobilities as suggested by the Hall conductivity measurements~\cite{akiba2}.
This complexity also makes difficult to find implication of the scattering parameter $r$ that evolves from -13 to -4 on approaching $P_{\rm c}$. 

\begin{figure}[tb]
\includegraphics[width=8cm]{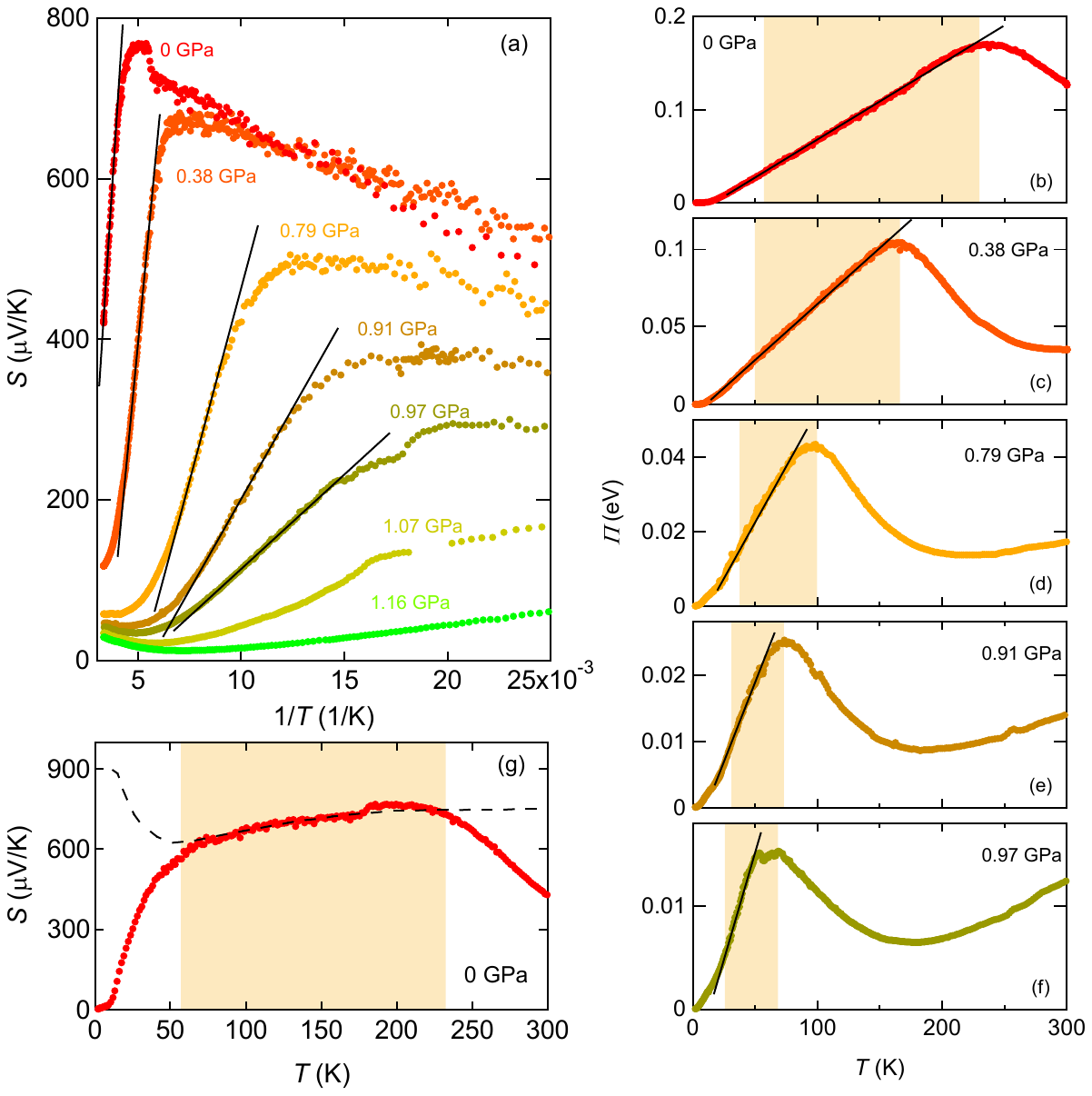}
\vspace*{-3cm} 
\caption{(a) Seebeck coefficient exhibits 1/$T$ dependence in the intrinsic regime as expected from the theory. 
Peltier coefficient also follows the expected $T$ dependence in the saturation regime, which is shaded by orange in the panels (b)-(f).
The panel (g) demonstrates that the Seebeck coefficient in the saturation regime shaded by orange is well described by the Pisarenko formula (Eq.~(9)). The calculated results are shown by a dashed line.}
\end{figure}

\begin{figure}[tb]
\includegraphics[width=8cm]{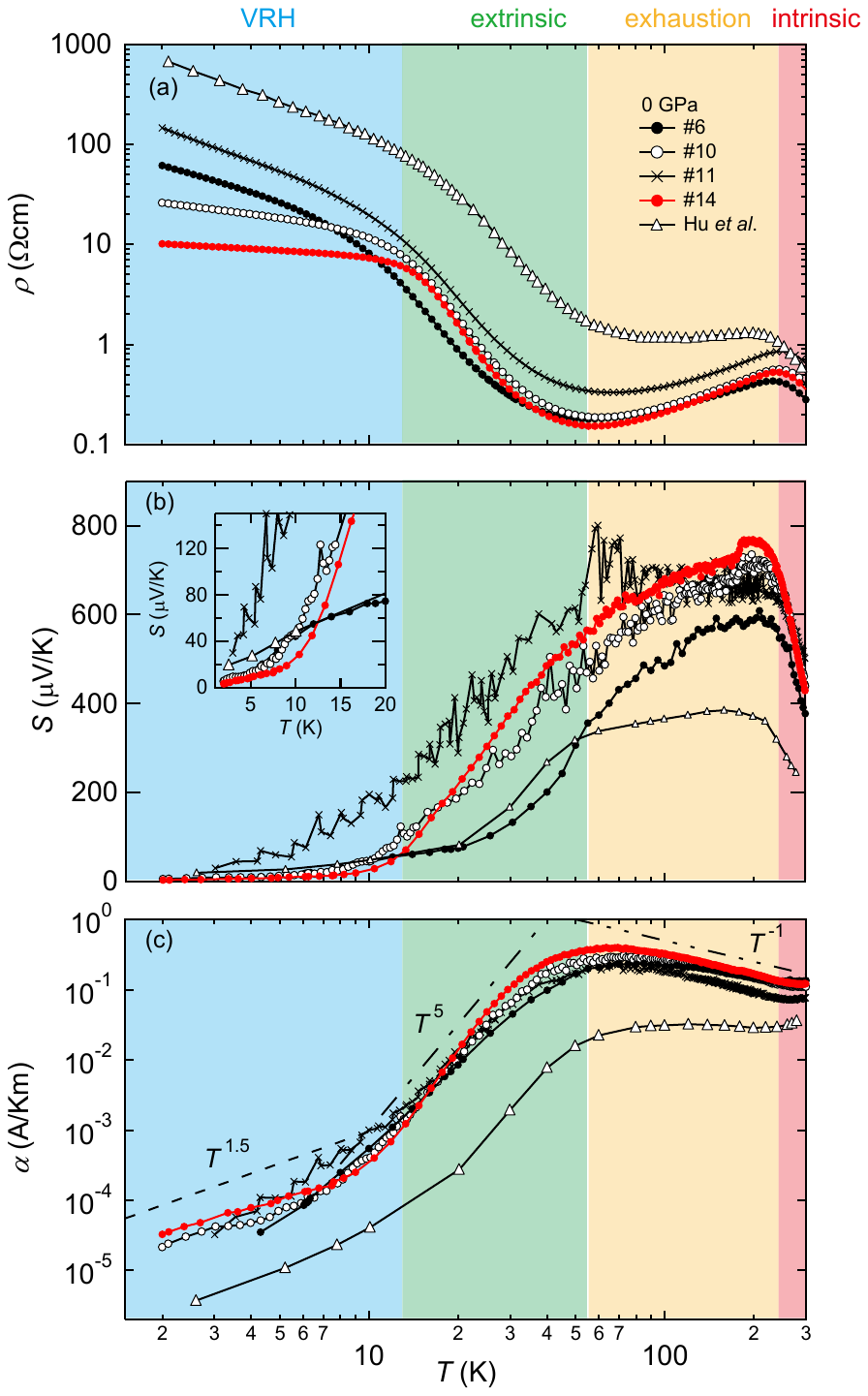}
\vspace*{-1.5cm} 
\caption{Temperature dependence of resistivity (a), Seebeck coefficient (b), and Peltier conductivity (c) for four different BP samples form the same batch together with the literature data~\cite{hu}. While $\rho$ and $S$ depend largely on the samples, $\alpha$ falls on an almost the same curve, which arrows us to access an intrinsic thermoelectric behavior of semiconductors.}
\end{figure}

An alternative estimation of $\Delta_{\rm vc}$ is provided by the Peltier coefficient $\Pi=ST$ (extracted using the Kelvin relation) which quantifies average thermal energy transported by carriers and
it corresponds to the activation energy $\Delta_{\rm vc}$.
Figures~8(b)-(f) depict temperature dependence of $\Pi$ at $P<P_{\rm c}$ (the saturation regime is shaded by orange).
$\Pi$ attains a maximum value at the lower temperature end of the intrinsic regime.
$\Delta_{\rm vc}$ extracted from the maximal $\Pi_{\rm max}$ as $e\Pi_{\rm max}=(\Delta_{\rm vc}/2T)T=\Delta_{\rm vc}/2$ is in good agreement with the one from resistivity, as seen in Fig.~5(a).
We note that this way of the gap estimation is compatible with the empirical method using the relation of $E_{\rm g}=2e|S_{\rm max}|T_{\rm max}$~\cite{gibbs}, where $E_{\rm g}$, $S_{\rm max}$, and $T_{\rm max}$ are the energy gap, the maximum Seebeck coefficient, and the temperature at which $S_{\rm max}$ occurs, respectively.
This yields $E_{\rm g}\sim$ 0.32 eV at ambient pressure, in good agreement with $\Delta_{\rm vc}\sim$ 0.3-0.4 eV extracted from resistivity and $\Pi$.

According to the Wilson-Sommerfeld treatment~\cite{geballe}, combining the equation of carrier density $n=A\exp(-\Delta_{\rm vc}/2k_{\rm B}T)$ ($A$ being constant) with $\Pi\propto\Delta_{\rm vc}$ gives the qualitative description of Peltier coefficient as,
\begin{equation}
\Pi\propto2k_{\rm B}T(\ln A-\ln n).
\end{equation}
Given that $n$ is nearly constant in the saturation regime, $\Pi$ is expected to vary proportional to temperature.
We show in figure 8(b)-(f) that $\Pi$ is indeed in $T$-linear in the saturation regime shaded by orange.
A quantitative account of the magnitude of Seebeck coefficient in the saturation regime is given by the Pisarenko formula~\cite{johnson,collignon} which is valid when the number of carriers is set by extrinsic dopants,
\begin{equation}
S=\cfrac{k_{\rm B}}{e}\left[2+r+\ln\left\{\cfrac{2}{n}\left(\cfrac{m^*k_{\rm B}T}{2\pi\hbar^2}\right)^{3/2}\right\}\right],
\end{equation}
where $m^*$ is the effective carrier mass. Since the dominance of phonon scattering between 50 K and 200 K is indicated by the Hall mobility experiments~\cite{akahama}, by setting $r$ = -1/2 and using $n$ extracted from the published data~\cite{akahama}, we obtained a fair agreement of the computed $S$ from Eq.~(9) to the experiments with $m^*$ = 0.45 $m_0$ in the saturation regime (see the dashed line in Fig.~8(g)). The deduced effective mass of $m^*$ = 0.45 $m_0$ reasonably agrees with the one extracted from the cyclotron resonance experiments~\cite{narita}.

Contrary to the activation behavior in resistivity, the Seebeck coefficient steeply drops upon entry to the extrinsic regime.
The similar behavior can be found in other semiconductors~\cite{ioffe,yoshikawa}.
While a satisfactory account of this contradiction is still missing, we infer that the presence of electron carriers whose mobility is comparable with that of hole carriers would cancel out the $1/T$ behavior because the Seebeck coefficient of two carrier system is expressed by Eq. (6).
The presence of electron carriers is backed by the fact that oxygen dopants induce n-doped BP~\cite{pei}.   

In the VRH regime, $S(T)$ varies linearly with temperature as shown  in the inset of Fig. 9(b). (See the data of sample \#14 from which the preceding data is obtained.)
Our observation is not agreement with what was seen in the BP nanoribbons in which $S(T)$ follows $T^{1/2}$ in the temperature range (between 50 K and 200 K) where resistivity obeys the VRH transport~\cite{liuBP}, and is in contrast with what was found in the doped Si~\cite{lakner} and the organic insulator~\cite{machidaPF} where $S(T)$ diverges on cooling. 
None of available theories predicting the vanishing VRH thermopower $S\sim T^{x}$ ($x$ = 1/2~\cite{zvyagin}, 3/4~\cite{yamamoto} for three dimensions) are compatible with our result.
We show here that the Peltier conductivity $\alpha=S/\rho$ rather than $S$ is useful to probe the intrinsic thermoelectric feature of semiconductors.
In Figs.~9(a)-(c), $\rho(T)$, $S(T)$, and $\alpha(T)$ of the several BP samples from the same batch together with the literature data~\cite{hu} are compared, respectively.
Clearly, $\rho(T)$ and $S(T)$ are highly sample dependent especially below the extrinsic regime because they are affected by the 
impurities.
By suppressing the extrinsic sample dependence in $\rho(T)$ and $S(T)$, however, all $\alpha(T)$ data fall into the same curve, allowing to address the intrinsic thermoelectric responses: $\alpha(T)$ varies with $T^{1.5}$ and $T^{5}$ in the VRH and extrinsic regimes, respectively.

Now, let us discuss the phonon drag contribution to the Seebeck coefficient in the semiconducting state.
In general, the Seebeck coefficient is composed of the diffusion term $S_{\rm dif}$ and the phonon drug term $S_{\rm ph}$,
\begin{equation}
S=S_{\rm dif}+S_{\rm ph}.
\end{equation}
The facts that the Seebeck coefficient obeys the activation behavior in the intrinsic regime and that the Pisarenko formula is valid in the saturation regime (both are features of the diffusive Seebeck response) indicate that a dominant role is played by the carrier diffusive in both regimes.
This means that in these regimes every response in the Seebeck coefficient is predominantly controlled by $\Delta_{\rm vc}$.
We do not exclude the phonon drag contribution below the extrinsic regime, but the lacking of quantitative explanation of the diffusive contribution prevents the further analysis. Notably, the gigantic phonon drag effect which arrows the Seebeck coefficient to attain as large as 6 mV/K below 50 K (the extrinsic region) is reported in germanium~\cite{geballe}.

\subsubsection*{semimetallic state}
The Seebeck effect in the semimetallic state is characterized by the presence of positive and negative peaks centered around 10 K and 30 K, respectively (red and blue triangles in Figs. 7(f)-(o)). The former persists up to $P^{*}$ where the dominant scatterers for carriers change from the ionized impurities to other electrons.
The latter appears as a dip above $P_{\rm c}$ and becomes a prominent negative peak above $P^{*}$.
To search for the origin of peaks, comparison to prototypical semimetals is instructive.
In semimetals, phonon drag effect provides an enhancement of Seebeck coefficient around a temperature where phonon scatter electrons most efficiently~\cite{issi}.
Such a condition is satisfied when typical phonon wave vector $q$ becomes comparable to twice of electron wave vector, $q\sim 2k_{\rm F}$. The effective Debye temperature $\Theta_{D}^{*}=2k_{\rm F}\bar{v}_{\rm s}\hbar/k_{\rm B}$ ($\bar{v}_{\rm s}$ being the average of phonon velocity), rather than the true Debye temperature, can be used to describe such a characteristic temperature. As shown in Table~1, the estimated $\Theta_{D}^{*}$ for Bi, Sb, and As is of the same order of magnitude as $T_{\rm max}$ where the peaks of Seebeck coefficient appear~\cite{uher,jaoui2,uher2}. Here, $k_{\rm F}$ for each system is evaluated from the carrier density~\cite{issi}.
Under the assumption that the calculated $\bar{v}_{\rm s}$ for the orthorhombic structure under ambient pressure~\cite{kaneta} is retained even under the pressure and with the knowledge of carrier density at 1.7 GPa~\cite{akiba2}, $\Theta_{D}^{*}$ for BP is estimated to be 12 K, which coincides with $T_{\rm max}\sim$ 10-30 K within experimental margin. 
This provides reasonable ground to argue that the phonon drag participates the formation of peaks.

\begin{table}[tb]
 \caption{Comparison of prominent semimetals with BP.}
  \centering
 \begin{tabular}{lcccc}
    \hline
    &Bi&Sb&As&BP\\
   \hline
$n$ (cm$^{-3}$)&2.7$ \times $10$^{17}$&3.7$ \times $10$^{19}$&2.0$ \times $10$^{20}$&2.9$ \times $10$^{17}$\\
$k_{\rm F}$ (nm$^{-1}$)&0.2&1.0&1.8&0.2\\
$\bar{v}_{\rm s}$ (km/s)&1.1&2.9&3.0&3.8\\
$\Theta_{\rm D}^*$ (K)&3.3&44&82&12\\
$T_{\rm{max}}$ (K)&3&10&30&10-30\\
Ref.&\cite{issi,uher}&\cite{issi,epstein,jaoui2}&\cite{issi,pace,uher2}&\cite{akiba2,kaneta}\\
   \hline
 \end{tabular}
\end{table}

\begin{figure}[tb]
\includegraphics[width=8.5cm]{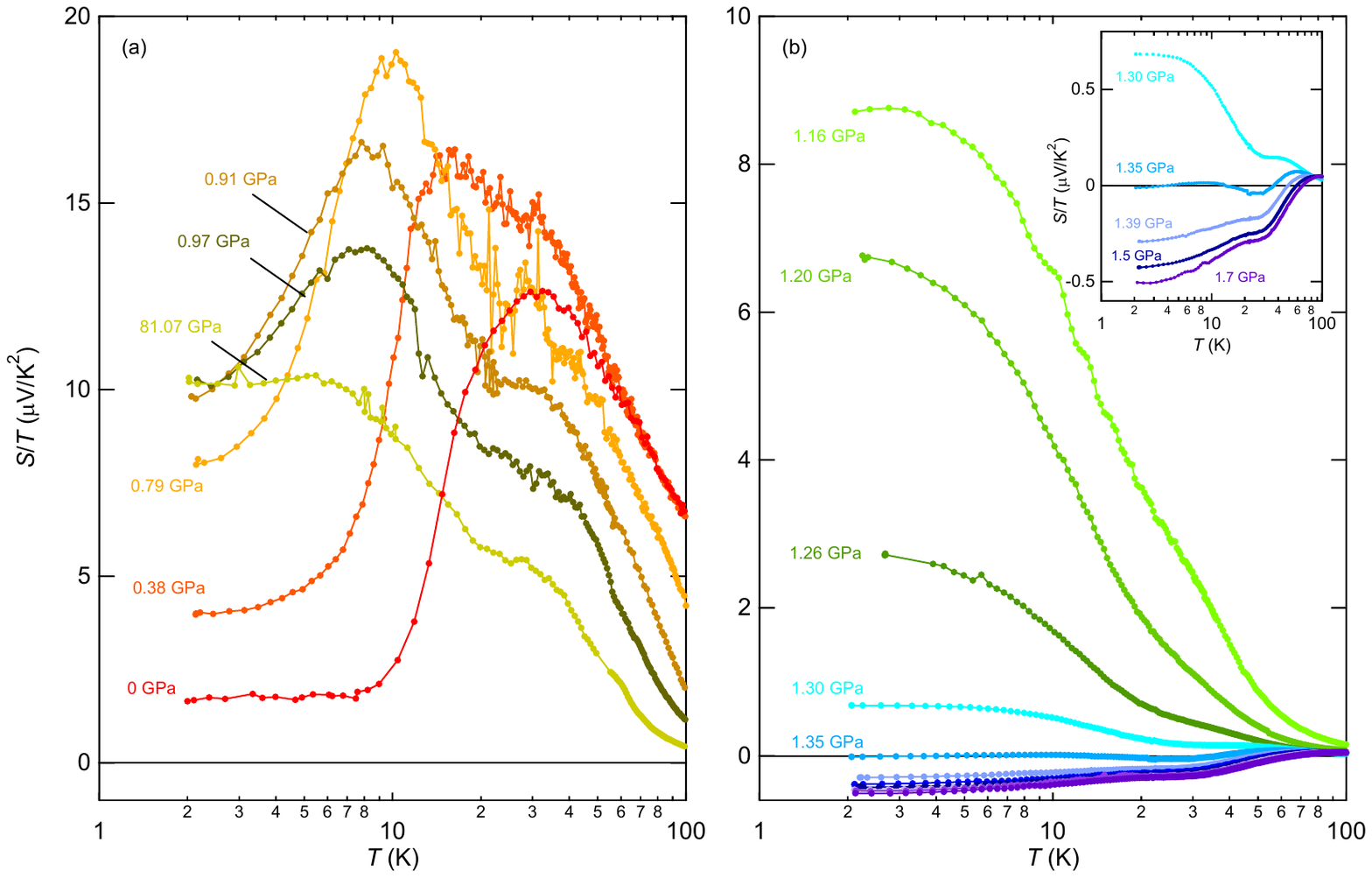}
\vspace*{-0.5cm} 
\caption{Seebeck coefficient divided by temperature $S/T$ as a function of temperature in the semiconducting state (a) and semimetallic state (b), respectively. The inset shows a zoom of the data at $P\geq$ 1.30 GPa. While $S/T$ is not constant at low temperature in the semiconducting state except at ambient pressure where the VRH transport governs the carrier conduction, it becomes constant in the semimetallic state as observed in ordinary metals.}
\end{figure}

\begin{figure}[tb]
\includegraphics[width=9cm]{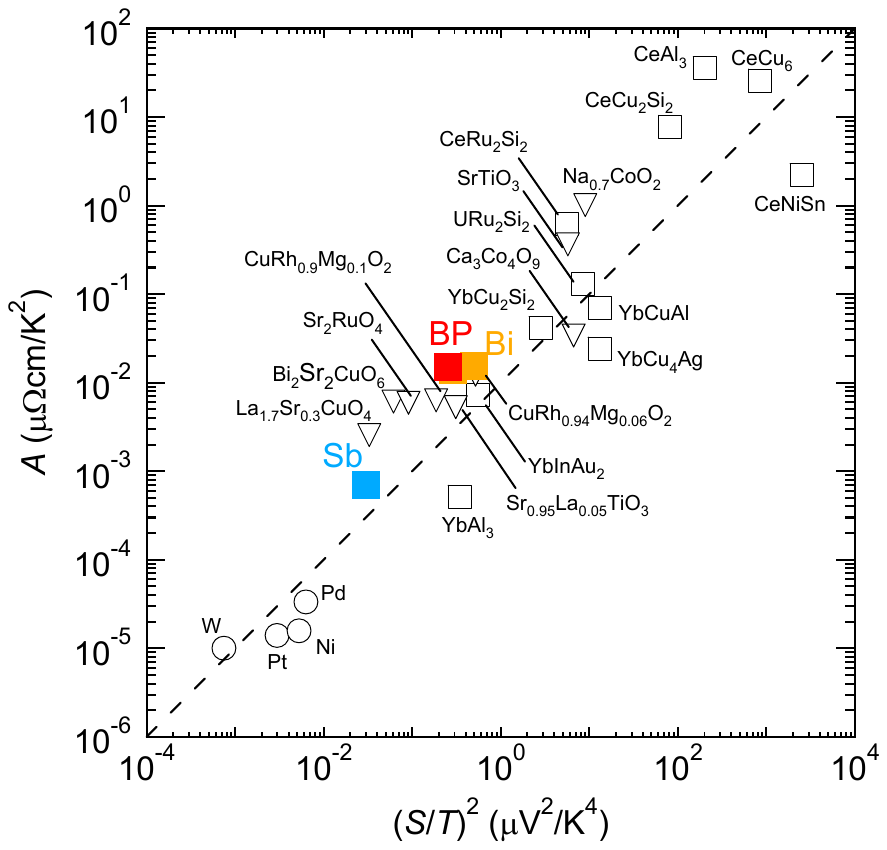}
\vspace*{-7cm} 
\caption{A plot of a prefactor of $T^2$-resistivity $A$ vs. $(S/T)^2$ for various materials. The data is adapted from Ref.~\cite{okazaki}. Note that not only prototypical semimetals of Bi and Sb, but also semimetallic BP (1.7 GPa) obey the universal relation between $A$ and $(S/T)^2$.}
\end{figure}

\begin{figure}[tb]
\includegraphics[width=9cm]{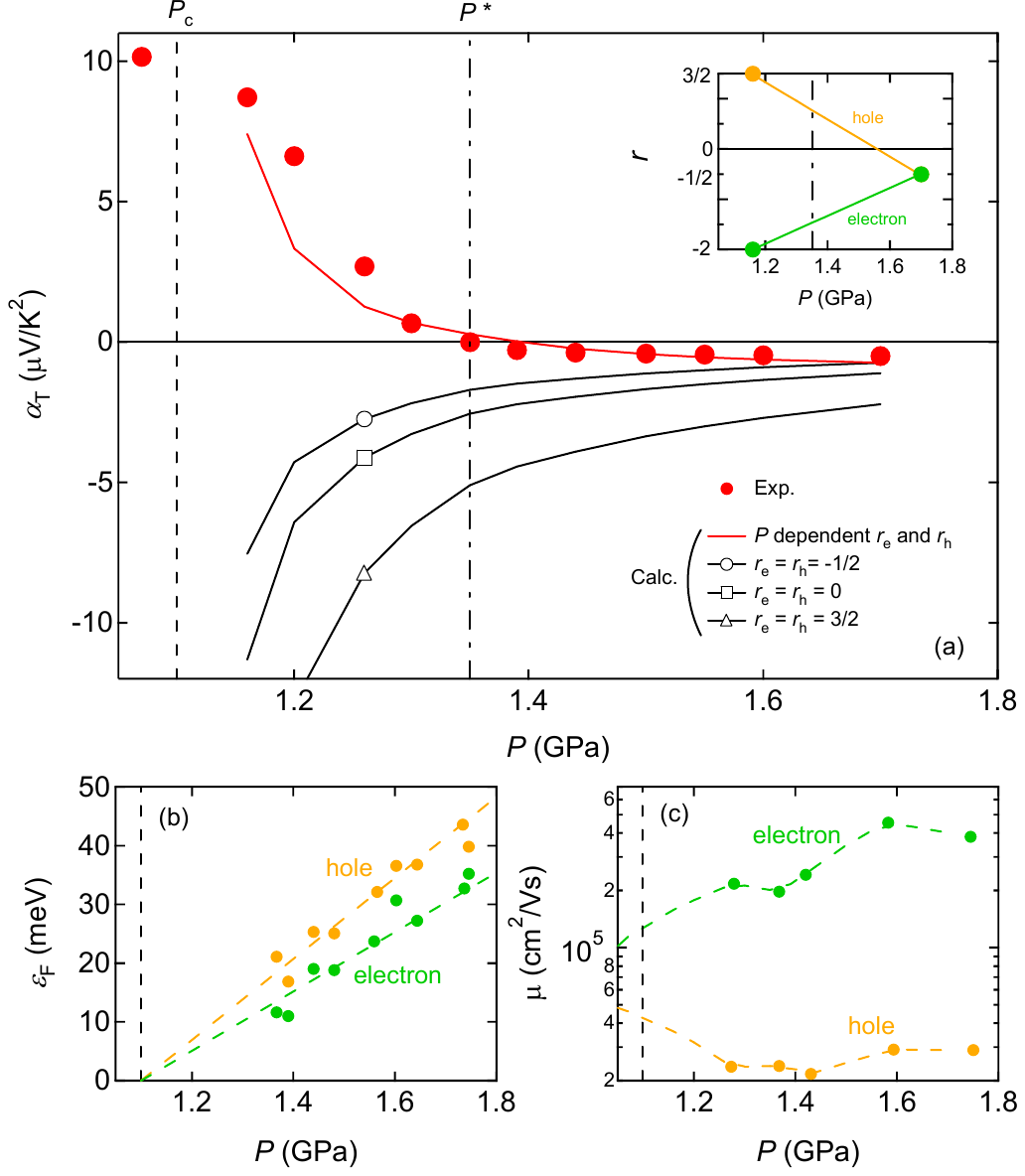}
\vspace*{-4cm} 
\caption{The $T$-linear term of Seebeck coefficient $\alpha_{\rm T}$ as a function of pressure.
The computed $\alpha_{\rm T}$ does not reproduce the experimental data when the scattering parameters $r_{\rm e,h}$ for electrons and holes are set as the same value.
An quantitative account of $\alpha_{\rm T}$ is possible when $r_{\rm e,h}$ varies with pressure as denoted by the solid red line.
Pressure dependence of the Fermi energy from Ref.~\cite{xiang} and mobility from Ref.~\cite{akiba2} for electrons and holes are shown in the panels (b) and (c), respectively.}
\end{figure}

A qualitative description of phonon drag effect is provided by Herring~\cite{herring} as,
\begin{equation}
S_{\rm ph}=\pm\Lambda\frac{k_{\rm B}}{e}\frac{m^*\bar{v}_{\rm s}^2}{k_{\rm B}T}\frac{\tau_{\rm ph}}{\tau_{\rm e}}.
\end{equation}
Here, $\Lambda$ is the momentum exchange rate between
phonons and electrons, and $\tau_{\rm ph}$ and $\tau_{\rm e}$ are phonon and electron scattering times.
According to this equation, a large phonon drag contribution to the Seebeck response is possible when $\tau_{\rm ph}\gg\tau_{\rm e}$~\cite{herring,jaouiInAs}.
Coming back to the giant $S_{\rm ph}$ in germanium, it is shown the large $\tau_{\rm ph}/\tau_{\rm e}$ ratio is a key to understand this phenomenon~\cite{geballe}.
For BP, the nearly ballistic heat conduction by phonons below 30 K~\cite{machidabp} enables the large $\tau_{\rm ph}$, while the ionized impurity scattering and the electron-electron scattering are responsible for the damping of $\tau_{\rm e}$.
The opposite sign of peaks implies the sign of carriers coupled with phonons is opposite. 
Given the Fermi surfaces constituting of two electron pockets and one hole pocket~\cite{akiba1,xiang,gong}, 
combining the spacial distribution of Fermi surfaces and phonon wave vectors of each branch would allow to pin down at which Fermi surface and at what temperature the favorable condition for the phonon drag effect ($q\sim 2k_{\rm F}$) is satisfied.
 
Upon cooling the sample down to low temperature, the phonon drag contribution dies out and diffusive contribution becomes significant, recovering the $T$-linear dependence in the Seebeck coefficient as expected from Eq.~(3). This is seen in figure 10(a)-(c), in which Seebeck coefficient divided by temperature $S/T$ under various pressures are shown as a function of temperature.
While $S/T$ in the semiconducting state ($P<$ 1.1 GPa) shows a striking temperature dependence with a concave curve and does not saturate except for under ambient pressure (Fig.~10(a)), the one in the semimetallic state exhibits a convex temperature dependence and eventually saturates to a constant value at low temperature (Fig. 10(b)). As shown in the inset of Fig.~10 (b), $S/T$ becomes constant but negative above $P^*$. The saturation behavior of $S/T$ suggests that the low-temperature Seebeck response is governed by carrier diffusion. Let us prove this conjecture by examining a universal relation between the prefactor of $T^2$ resistivity, $A$, and the $T$-linear term of Seebeck coefficient~\cite{okazaki}.
It is seen from Fig.~11 that not only prominent semimetals of Bi and Sb, but also BP for which $A$ and $S/T$ are evaluated under 1.7 GPa fall into the universal line, providing evidence that the thermoelectric response is
purely diffusive at low temperature.

Having established the diffusive response in the semimetallic state, let us scrutinize the response of the $T$-linear term of the Seebeck coefficient $\alpha_{\rm T}$ against the pressure.
As shown in Fig.~12(a), $\alpha_{\rm T}$ falls above $P_{\rm c}$ and changes the sign across $P^*$.
By combining Eq.~(3) and (6), one gets the expression of $\alpha_{\rm T}$ for semimetals with equal number of electron and hole carrier density ($n_{\rm e}=n_{\rm h}$),
\begin{equation}
\alpha_{\rm T}=-\frac{\pi^2}{3}\frac{k_{\rm B}^2}{e(\mu_{\rm e}+\mu_{\rm h})}\biggl\{\frac{\mu_{\rm e}}{\epsilon_{\rm F}^{\rm e}}\biggl(r_{\rm e}+\frac{3}{2}\biggr)-\frac{\mu_{\rm h}}{\epsilon_{\rm F}^{\rm h}}\biggl(r_{\rm h}+\frac{3}{2}\biggr)\biggr\},
\end{equation} 
where $\mu_{\rm e}$ ($\mu_{\rm h}$),  $\epsilon_{\rm F}^{\rm e}$ ($\epsilon_{\rm F}^{\rm h}$), and $r_{\rm e}$ ($r_{\rm h}$) are mobility, Fermi energy, and the exponent of energy dependence of scattering rate of electrons (holes), respectively.
To compute the pressure dependence of $\alpha_{\rm T}$, one needs $\mu_{\rm e,h}$ and $\epsilon_{\rm F}^{\rm e,h}$ as a function of pressure which are taken from Ref.~\cite{akiba2} and \cite{xiang}, and shown in Fig.~12(b) and (c), respectively. Since the data for $\epsilon_{\rm F}^{\rm e,h}$ is interrupted around 1.3 GPa, $\epsilon_{\rm F}^{\rm e,h}$ is assumed to vanish at $P_{\rm c}$ as shown by the dashed line in Fig.~12(b).
Due to the higher mobility and the lower Fermi energy of electrons, the computed $\alpha_{\rm T}$ is always dominated by electrons ($\alpha_{\rm T}$ is always negative) and does not reproduce the experimental data as long as $r_{\rm e,h}$ is set to be the same value and pressure-independent. 
The cases of $r_{\rm e}=r_{\rm h}=-1/2$, $r_{\rm e}=r_{\rm h}=0$, and $r_{\rm e}=r_{\rm h}=3/2$ are shown in Fig.~12(a). Nevertheless, the better agreement with the experiment is obtained for $P>P^*$ with $r_{\rm e}=r_{\rm h}=-1/2$. This corresponds to an energy-independent mean-free-path, which is a good description for electron scattering even in the presence of electronic correlations.
An elaborate agreement including the sign change in $\alpha_{\rm T}$ is achieved with the assumption that $r_{\rm e}$ and $r_{\rm h}$ evolves differently with pressure.
For the case where $r_{\rm h}$ and $r_{\rm e}$ start to changes from 3/2 and -2, respectively, and merge to be -1/2 at the end (see the inset of Fig.~12(a)), not only the magnitude of $\alpha_{\rm T}$ but also the sign reversal at $P^*$ can be reproduced by the computations as show by the solid red line in Fig.~12(a). 
Such an evolution of $r_{\rm h}$ with pressure is in good accordance with our observation in resistivity that the dominant scattering process affecting carriers changes from ionized impurity scattering ($r_{\rm h}=3/2$) to electron-electron scattering ($r_{\rm h}=-1/2$) across $P^*$. While the justification of $r_{\rm e}$ = -2 near $P_{\rm c}$ remains a challenge, an equal influence of ionized impurity scattering for electrons and holes that was theoretically proposed in Ref.~\cite{blatt} would support our analysis.
What we emphasize here is that in addition to the thermodynamic quantity like $\epsilon_{\rm F}$, how and by what are carriers scattered is a key ingredient that determines the magnitude as well as the sign of Seebeck coefficient.

\section*{IV. SUMMARY}
In summary, our measurements on the Seebeck coefficient of black phosphorus under pressure have provided several piece of information that promotes our understanding of thermoelectricity in solids.
We showed that the Seebeck coefficient follows the activation behavior in the intrinsic regime and yields the gaps which are reasonably in agreement with what was determined by other experiments.
The validity of the Pisarenko formula in the saturation regime is documented with effective mass comparable with the one obtained by the cyclotron resonant experiments.
We resolved the vanishing fate for the Seebeck coefficient with $S\sim T$ in the VRH regime whereas its $T$-linear slope is highly sample dependent.
The independence of the Peltier conductivity $\alpha$ on the different samples paves a way to access the intrinsic thermoelectric properties even in the extrinsic semiconductors.

In the semimetallic state, we found that the increasing of carrier density induces the change in dominant scattering mechanism from the ionized impurity scattering to electron-electron scattering.
Both mechanism participate the enhancement of phonon drag effect at high temperature, but the opposite sign of phonon drag peaks imply distinct type of carriers are dragged by phonons. 
The intimate link between the change in dominant scatterers for carriers and the sign reversal of low temperature Seebeck coefficient that is purely dominated by carrier diffusion explicitly points to the crucial role of carrier scattering in determining not only the magnitude but also the sign of Seebeck coefficient.  

\section{ACKNOWLEDGMENTS}
We acknowledge discussions with Kamran Behnia. 
This work was supported by Grants-in-Aid for Scientific Research (KAKENHI Grants Nos. JP16K05435, JP17KK0088, and JP19H01840).

\section{APPENDIX: PRESSURE EFFECT ON TYPE E THERMOCOUPLE}
Pressure effect on the type E thermocouple was investigated by using the setup shown in Fig.~13. The junction of thermocouple was glued by silver paste (DuPont 4922N) on one side of a Cernox thermometer (CX-1050-HT). A chip resistor was attached on the other side of the thermometer, which serves the heat current along the thermocouple.
The other end of thermocouple was thermally anchored to the body of pressure cell and connected to the copper wires for the thermoelectric vlotage ($V_{\rm th}$) measurement.
The temperature gradient generated along the thermocouple was measured by using two Cernox thermometers.
One was at the thermocouple junction as noted above and monitored the temperature ($T_{\rm in}$) at the junction (hot end of the thermocouple) inside the pressure cell. 
The other one was placed on the body of pressure cell and measured the temperature ($T_{\rm out}$) at the cold end of thermocouple. The latter was always under ambient pressure. 
An external pressure was applied by using a NiCrAl-BeCu hybrid piston cylinder. 
Daphne oil 7373 was used as a pressure transmitting medium.
The applied pressure was estimated by monitoring the superconducting transition temperature of Pb.
\begin{figure}[b]
\includegraphics[width=9cm]{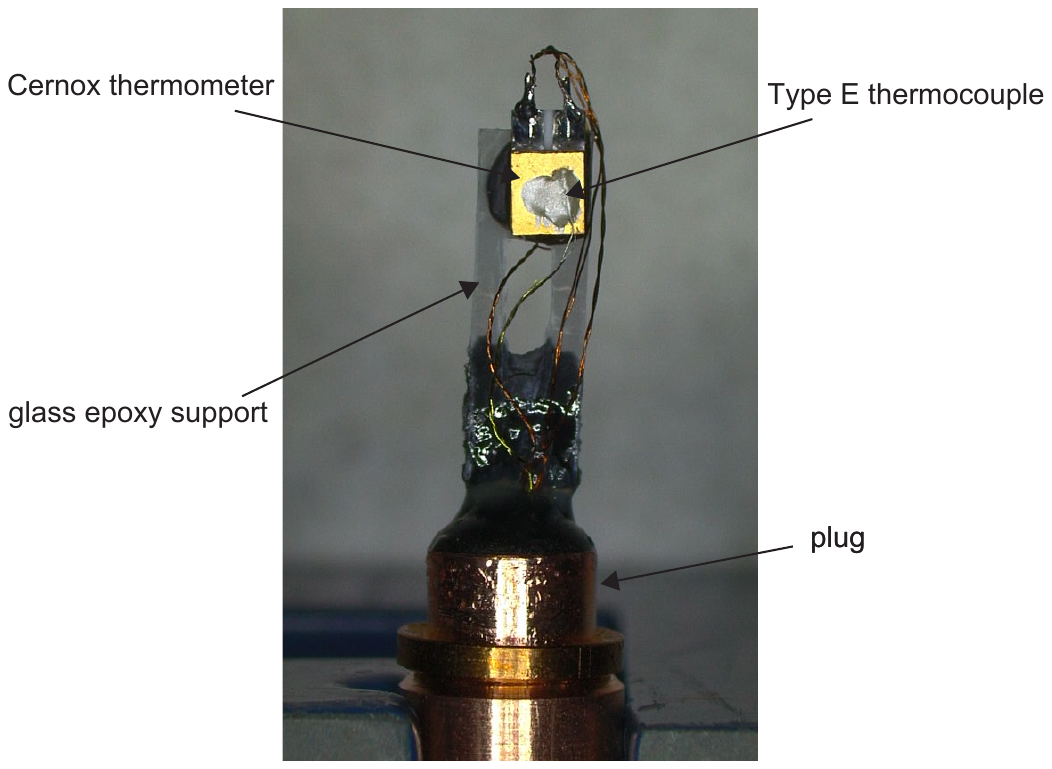}
\vspace*{-7cm} 
\caption{A photo of setup for calibration of the type E thermocouple under pressure.}
\end{figure}
\begin{figure}[tb]
\includegraphics[width=9cm]{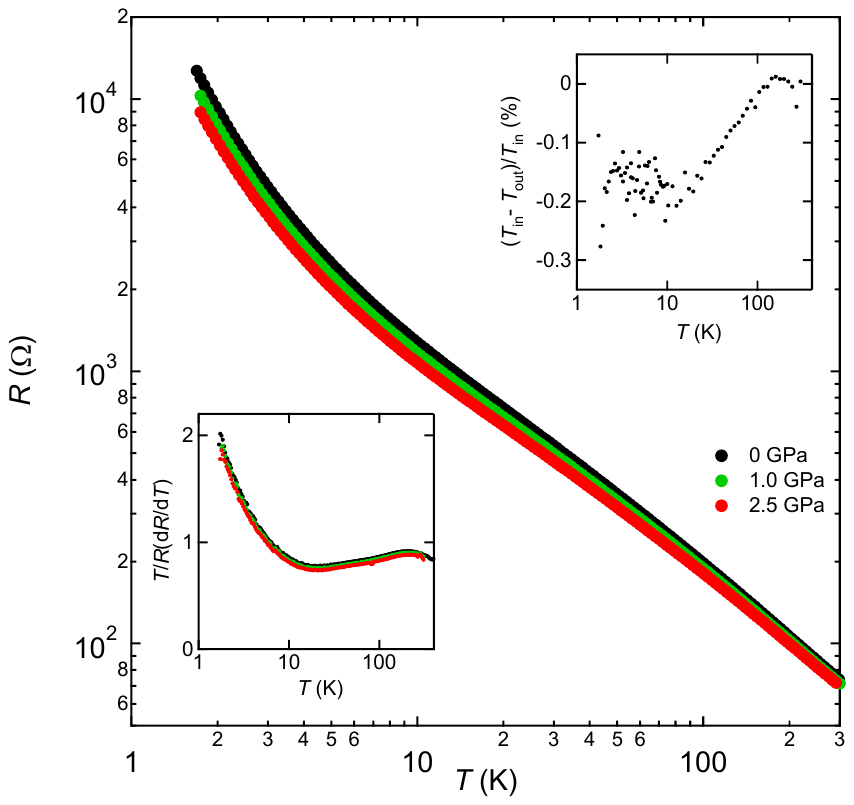}
\vspace*{-7cm} 
\caption{Temperature dependence of Cernox thermometer at different applied pressures. Upper inset demonstrates that temperature difference between Cernox thermometers placed inside and outside of the pressure cell is less that 0.3 \% at ambient pressure. Lower inset: dimensionless sensitivity of thermometer $(T/R)(dR/dT)$ at each pressure.}
\end{figure}

We first ensured that there is no appreciable temperature difference between the two thermometers located inside and outside the pressure cell at ambient pressure without injecting the heat current. The upper inset of Fig.~14 shows that the temperature difference between the thermometers $(T_{\rm in}-T_{\rm out})/T_{\rm in}$, which is less than 0.3 \% in the whole temperature range measured. 
Subsequently, the thermometer inside of the cell was calibrated under pressure against the thermometer outside of the cell. The results are shown in the main panel of Fig.~14.
While the resistance of thermometer slightly decreases with pressure, its dimensionless sensitivity $(T/R)dR/dT$ is not degraded by the pressure as shown in the lower inset of Fig.~14.

\begin{figure}[tb]
\includegraphics[width=9cm]{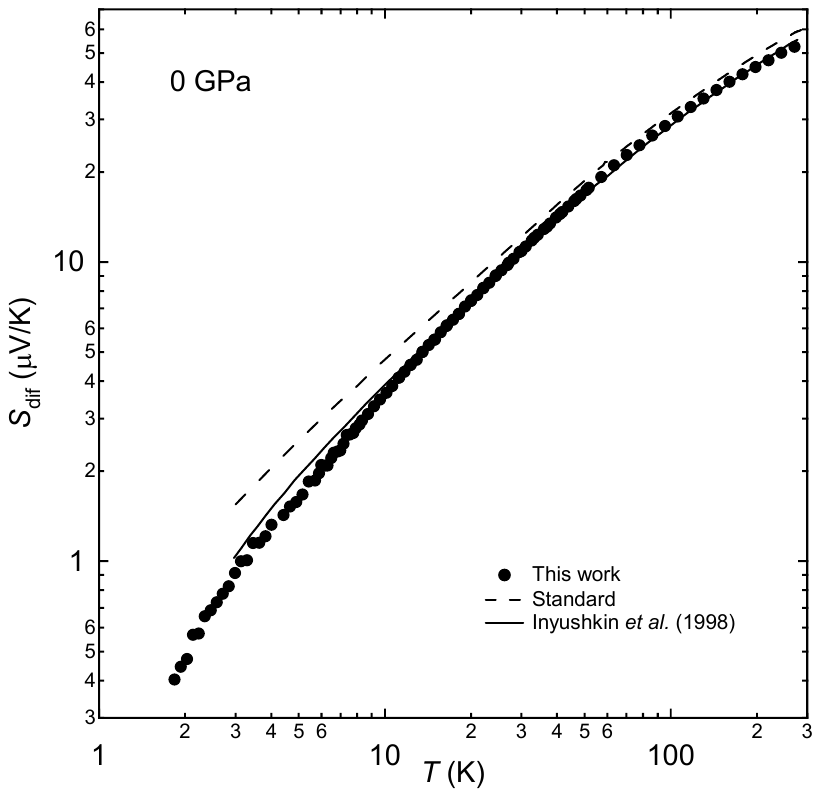}
\vspace*{-7cm} 
\caption{Temperature dependence of differential Seebeck coefficient $S_{\rm dif}$ of type E thermocouple at ambient pressure. For comparison, the standard data~\cite{powell} and the literature data~\cite{inyushkin} are also shown.}
\end{figure}
\begin{figure}[tb]
\includegraphics[width=9cm]{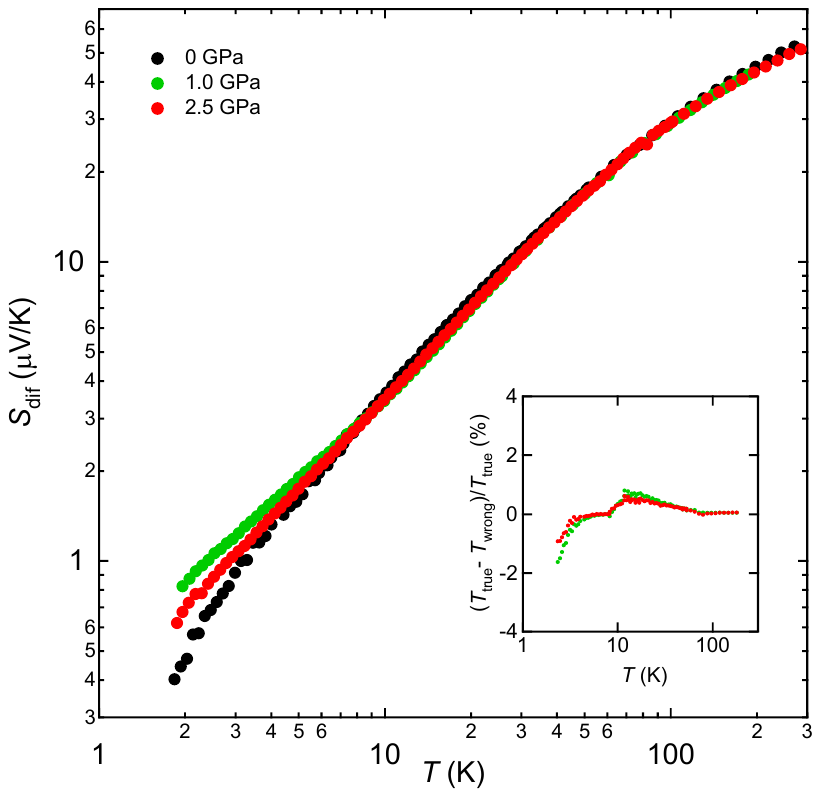}
\vspace*{-7cm} 
\caption{Temperature dependence of the differential Seebeck coefficient $S_{\rm dif}$ of the type E thermocouple measured under three different pressures. Inset shows the correction on temperature at the thermocouple junction.}
\end{figure}

Figure 15 shows temperature dependence of the differential Seebeck coefficient $S_{\rm dif}$ of the type E thermocouple at ambient pressure.
For comparison, the standard data~\cite{powell} and the literature data~\cite{inyushkin} are also shown. As seen from the figure, our data well coincides the one from Ref.~\cite{inyushkin} and deviates from the standard data especially at low temperature.
The discrepancy may be due to different amount of impurities contained in the thermocouple wires made by different manufacturers; our constantan and chromel wires are from Goodfellow, and the constantan wire from Goodfellow and the chromel wire from Hoskins Manufacturing were used in Ref.~\cite{inyushkin} (manufacturers are not specified for the standard data). 

Figure 16 shows temperature dependence of $S_{\rm dif}$ measured under three different pressures. Significant dependence of $S_{\rm dif}$ on the pressure is observed only below 10 K. $S_{\rm dif}$ increases at most 80 \% than that of ambient pressure at 1 GPa and 2 K. As shown in the inset of Fig. 16, the relative change of $S_{\rm dif}$ with pressure results in less than 2 \% of the correction on temperature at the thermocouple junction estimated as $(T_{\rm true}-T_{\rm wrong})/T_{\rm true}$, where $T_{\rm true}=-V_{\rm th}/S(P)+T_{\rm out}$ and $T_{\rm wrong}=-V_{\rm th}/S$(0 GPa) + $T_{\rm out}$. This small correction led us to conclude that the pressure effect on the type E thermocouple is neglectable up to 2.5 GPa.

\end{document}